\documentclass[titlepage,12pt,reqno]{article}
\usepackage{amsmath,amssymb,amsfonts,epsf,graphics}
\allowdisplaybreaks

\begin{document} 


\begin{titlepage}

\centerline{\LARGE \bf Nonequilibrium dynamics of the $\sigma$-model modes} 
\centerline{\LARGE \bf on the de Sitter space}
\vskip 1.5cm
\centerline{ \bf Ion V. Vancea }
\vskip 0.5cm
\centerline{\sl Grupo de F{\'{\i}}sica Te\'{o}rica e Matem\'{a}tica F\'{\i}sica, Departamento de F\'{\i}sica}
\centerline{\sl Universidade Federal Rural do Rio de Janeiro}
\centerline{\sl Cx. Postal 23851, BR 465 Km 7, 23890-000 Serop\'{e}dica - RJ,
Brasil}
\centerline{
\texttt{\small ionvancea@ufrrj.br} 
}

\vspace{0.5cm}

\centerline{19 January 2017}

\vskip 1.4cm
\centerline{\large\bf Abstract}

The two-dimensional $\sigma$-model with the de Sitter target space has a local canonical description in the north pole diamond of the Penrose diagram in the cosmological gauge. The left and right moving modes on the embedded base space with the topology of a cylinder are entangled among themselves and interact with the time-dependent components of the metric of the de Sitter space. Firstly we address the issue of the existence of the untangled oscillator representation and the description of the nonequilibrium dynamics of the untangled modes. We show that the untangled oscillators can be obtained from the entangled operators by applying a set of Bogoliubov transformations that satisfy a set of constraints that result from the requirement that the partial evolution generator be diagonal. Secondly, we determine the nonequilibrium dynamics of the untangled modes in the Non-Equilibrium Thermo Field Dynamics formalism. In this setting, the thermal modes are represented as thermal doublet oscillators that satisfy partial evolution equations of Heisenberg-type. We use these equations to compute the local free one-body propagator of an arbitrary mode between two times. Thirdly, we discuss the field representation of the thermal modes. We show that there is a set of thermal doublet fields that satisfy the equal time canonical commutation relations, are solutions to the $\sigma$-model equations of motion and can be decomposed in terms of thermal doublet oscillators. Finally, we construct a local partial evolution functional of Hamilton-like form for the thermal doublet fields. 

\vskip 0.7cm 


\noindent 

\end{titlepage}


\section{Introduction}

The two-dimensional non-linear $\sigma$-model with the de Sitter target space describes a non-trivial class of field theories of which quantum dynamics is still largely unknown. Since the $\sigma$-models have proved themselves instrumental to the description of the string theory on several flat and curved space-time manifolds, it is likely that, by studying them on the de Sitter space we will gain important insight into the structure of the string theory compactified to the four-dimensional expanding space-time and, more generally, on the time-dependent hyperbolic manifolds. 

A considerable hindrance to the formulation of the quantum field theories on the de Sitter space has been the global structure of the latter that has made unclear the physical interpretation of different mathematical objects, e. g. the generators of the de Sitter group, and the application of fundamental quantum concepts, e. g. the $S$-matrix. 
To some degree, it is possible to apply the Wightman two-point function formalism to non-interacting fields in the principal series representations of the $SO(1,4)$ group used to define the massive de Sitter fields and the Gupta-Bleuler method to the massless fields in the discrete series representations \cite{Gazeau:2006gq}. However, there are still complications related to the causal horizons, the background temperature and the instability of the system to the vacuum decay. Due to these particularities, it is extremely difficult to apply consistently even the non-covariant quantization methods to the field theories on the de Sitter space. 

In contrast to other field theories, the two-dimensional $\sigma$-model is naturally localized on its base space which can be embedded inside a local causal region of the de Sitter space that is accessible to a single observer when the typical scale of the base space is much smaller than that of the target space. This localization usually implies that some local symmetries of the model have been fixed or linearized. In these conditions, the standard quantization methods can be employed\footnote{The obvious drawback of this approach is that the quantization is both local and non-covariant. 
Yet, in the absence of a covariant method, this is the only approach that has provided any information on the quantum properties of the $\sigma$-model on the de Sitter space so far.}. Moreover, if the base space is considered to be a closed two dimensional surface, then the corresponding $\sigma$-model contains an inner interaction of modes propagating in opposite directions on the base space. This interaction can be non-trivial in the de Sitter space even in a linearized theory.

There has been some work done recently along this line of investigation in the framework of the canonical quantization in time-dependent gauges that attempted to interpret the $\sigma$-models as tree level bosonic string theories\footnote{The de Sitter space-time is not a proper string background as it is not a solution of the string $\beta$-function equations.}.  The results obtained so far include solving the equation of state for the perturbations around the center of mass of different string configurations in the conformal gauge, the construction of the corresponding time-dependent Fock space and the calculation of semi-classical quantities such as the mass of states and the maximum number of a single string excitation \cite{deVega:1987veo,deVega:1992yz,DeVega:1992xc,deVega:1994yz,RamonMedrano:1999gm,Bouchareb:2005ck}. The same quantities were derived for the little strings in an arbitrary time-dependent gauge in \cite{Li:2007gf}. The divergences of the Nambu-Goto model were discussed in the path integral formulation in \cite{Viswanathan:1996yg} and the back-reaction in the classical theory was analysed in \cite{Bozhilov:2001kw}. 

In a very recent paper, the $\sigma$-model with a two-dimensional cylindrical base space has been quantized in a new time-dependent gauge, called the cosmological gauge, that is natural to the model in the sense that the equal time commutation relations of the fields and the canonical commutation relations of their oscillating modes are compatible with each other in this gauge \cite{Vancea:2016tkt}. It was shown there that the left and right quantum excitations 
display a time-dependent entanglement generated by the partial evolution map that acts between the states at two instances of time and has the $su(1,1)^{\infty}$ symmetry\footnote{States with $SU(1,1)$ symmetry in the early universe have been discussed recently in \cite{Brahma:2016wvk}.}. The partial evolution map can be expressed as an unitary evolution operator generated locally by a hermitian Hamiltonian-like operator in the north pole diamond of the Penrose diagram of the de Sitter space. But in contrast with the flat space-time, the oscillating modes are quenches of the generator of the partial evolution map and not eigenvectors of it. That raises the question whether there is a representation of the oscillating modes in which the generator is diagonal and the oscillators are untangled. Finding such representation is a necessary step to addressing the issue of the physical properties of the $\sigma$-model excitations. 
 
In this paper, we propose a simple solution to the problem of the untangled oscillator representation by constructing a set of linear time-dependent Bogoliubov transformations that diagonalize the local Hamiltonian-like generator of the partial evolution map. We show that the real time-dependent functions that parametrize the Bogoliubov transformations satisfy a set of constraint equations that represent the diagonalization condition. Also, we determine the differential Heisenberg-like equations that describe the partial evolution of the untangled oscillators. The thermal effects in a quantum system in a curved space-time can be considered among the most fundamental physical properties. The untangled oscillator representation allows one to define and analyse the thermalization of the excitations of the $\sigma$-model in the de Sitter space by applying canonical methods. The fact that the frequencies depend on time suggests that the thermal dynamics of the oscillators should take place in nonequilibrium. We examine the thermalization of the untangled oscillators in the Non-Equilibrium Thermo Field Dynamics (NETFD) formalism in which the degrees of freedom of the thermal oscillators are most naturally represented by thermal doublets \cite{Umezawa:1993yq,Mizutani:2011rt,Mizutani:2011jy}. That allows us to determine the local nonequilibrium dynamics of the untangled modes generated by the partial evolution operator and calculate the propagator of an arbitrary thermal mode between two different instants of time. In curved space-times, the compatibility between the particle excitations and the field description is an important problem to be analysed. In the present case, it is important to know whether the thermal modes correspond to a thermal field theory of the $\sigma$-model. We answer this question in affirmative by constructing the thermal doublet fields associated to the $\sigma$-model we started with and their canonically conjugate momenta. These fields have the following fundamental properties: their oscillating modes are the thermal doublet oscillators, they are solutions to the $\sigma$-model equations of motion in the cosmological gauge and the canonically conjugate fields satisfy the standard equal time commutation relations. The partial evolution of the thermal fields is characterized by differential equations that can be interpreted as canonical extended Heisenberg equations. This observation allows us to derive the classical partial evolution functional of Hamiltonian-type for the thermal doublet fields.  

The paper is organized as follows. In Section 2 we review the results from \cite{Vancea:2016tkt} on the quantization of the $\sigma$-model in the cosmological gauge that are necessary to obtain the untangled oscillator representation. Then 
we define the Bogoliubov transformations between the entangled and untangled oscillators and derive the constraints satisfied by the Bogoliubov parameters. From the diagonal generator of the partial evolution, we deduce the partial evolution equations of the untangled oscillators. In Section 3 we describe the thermalization of the untangled oscillators in the NETFD approach. We formulate the thermal degrees of freedom as thermal doublet oscillators. Then we give a second set of Bogoliubov transformations that map the untangled oscillators into two different representations: the time-independent and the time-dependent pseudo-particle representations, respectively, which are needed to define the vacuum in which the physical quantities should be computed and to determine the dynamics of the thermal oscillators. 
These are used to calculate the propagator of an arbitrary thermal doublet mode between two different values of time. In Section 4 we obtain the thermal doublet fields associated to the thermal doublet oscillators. These form of this fields 
satisfy the $\sigma$-model equations of motion and we shown that they satisfy the standard equal time commutation relations. Also, the partial evolution functional corresponding to this structure is derived. In the last section we discuss the results. In the Appendix can be found a list of the fundamental axioms of the NETFD that are used throughout this work.

\section{Quantum $\sigma$-model in the cosmological gauge}

In this section we review some results on the quantization of the two-dimensional $\sigma$-model with the de Sitter target space obtained in \cite{Vancea:2016tkt} that are relevant for the construction of the untangled oscillator representation. Then we define the Bogoliubov transformations between the entangled and untangled oscillator representations and determine the constraints that must be satisfied by the Bogoliubov parameters and the partial evolution of the untangled oscillators.

\subsection{Canonical $\sigma$-model in the cosmological gauge}  

The $\sigma$-model of interest is defined by the embedding of the two-dimensional cylinder $\Sigma^{1,1} = S^{1} \times I$, $I \subset \mathbb{R}$ into the north pole diamond of the Penrose diagram of the four-dimensional de Sitter target space denoted by $dS_4$. The base space $\Sigma^{1,1}$ is endowed with the metric $ h_{ab}(\tau, \sigma)$ and the comoving frame is chosen in $dS_4$ in which the metric takes the following form
\begin{equation}
ds^2 = - dt^2 + e^{2Ht}\delta_{ij} dx^i dx^j .
\label{dS4-metric}
\end{equation}
The spatial and temporal coordinates on the base space are denoted by $\sigma \in [0,2 \pi]$ and $\tau \in I$, respectively. The embedding is characterized by the orientation $t = \tau$, where $t$ is the parameter of the local time-like Killing vector in the north pole diamond of $dS_4$. The model is invariant under the two-dimensional Weyl symmetry and the reparametrization of the variables $\tau$ and $\sigma$. Then one can pick up the cosmological gauge which is defined as the following time-dependent gauge of the Weyl symmetry 
\begin{equation}
h^{-1}_{00} (t = \tau, \sigma)= h_{11} (t = \tau, \sigma)= e^{-2Ht}, 
\hspace{0.5cm}
h_{01} (t = \tau, \sigma) = 0.
\label{cosmological-gauge}
\end{equation}
Here, $H$ denotes the Hubble's constant. In the cosmological gauge, the canonical commutation relations of the
quantum oscillators and the equal time commutation relations of the corresponding canonical fields 
are compatible to each other. This statement is not true for an arbitrary time-dependent gauge \cite{Vancea:2016tkt}. 
In this setting, the action of the $\sigma$-model takes the following form
\begin{equation}
S[x] =  \frac{1}{4 \pi} \int dt d\sigma
\left[ 
- e^{-2Ht} + \left( \partial_t x^{i} (t,\sigma) \right)^2
- e^{2Ht} \left( \partial_{\sigma} x^{i} (t,\sigma) \right)^2
\right],
\label{fixed-action-cosmo-gauge}
\end{equation}
where $x^{i} = x^1, x^2, x^3$ are the spatial direction transverse to the local time-like Killing vector in $dS_4$ that span the transverse section with the $SO(3)$ symmetry. The equations of motion that correspond to the action $S[x]$ take the following form
\begin{equation}
\partial^{2}_{t} x^i (t,\sigma) - 
 e^{4Ht}\partial^{2}_{\sigma}
x^i (t,\sigma) = 0.
\label{eq-motion-class}
\end{equation} 
The general solution of the equation (\ref{eq-motion-class}) and its conjugate momenta have the following linear decomposition in terms of the Hankel functions \cite{Olver:2010} that reflect the symmetry of the embedded base space
\begin{align}
x^{i}\left( t,\sigma \right) & = x_{0}^{i}+p_{0}^{i}t
+\frac{i}{2}
\sqrt{\frac{\pi}{H}}\sum _{m>0}\left[ \alpha _{m}^{i}e^{im\sigma }+\beta _{m}^{i}e^{-im\sigma }\right] H_{0}^{(2)}\left( z_{m}\right)
\nonumber
\\
& -\dfrac {i} {2}\sqrt {\dfrac {\pi } {H}}\sum _{m>0}\left[ \begin{matrix} \alpha _{m}^{i\dagger}e^{-im\sigma }& +\beta _{m}^{i\dagger}e^{im\sigma }\end{matrix} \right] H_{0}^{(1)}\left( z_{m}\right)
,
\label{sol-x-herm}
\\
\pi^{i}\left( t,\sigma \right) & = \frac{1}{2\pi} p_{0}^{i}
+\frac{i}{2}
\sqrt{\frac{H}{\pi}}\sum _{m>0}\left[ \alpha _{m}^{i}e^{im\sigma }+\beta _{m}^{i}e^{-im\sigma }\right] 
z_m
\dfrac{d H_{0}^{(2)}\left( z_{m}\right)}{dz_m}
\nonumber
\\
& -\dfrac {i} {2}\sqrt {\dfrac {H } {\pi}}\sum _{m>0}\left[ \begin{matrix} \alpha _{m}^{i\dagger}e^{-im\sigma }& +\beta _{m}^{i\dagger}e^{im\sigma }\end{matrix} \right] z_m \dfrac{dH_{0}^{(1)}\left( z_{m}\right)}{d z_m}
,
\label{sol-p-herm}
\end{align}
where the conjugate momenta are defined by 
\begin{equation}
\pi^{i} (t,\sigma)= \frac{1}{2 \pi} \partial_{t} x^{i} (t,\sigma).
\label{mom-def}
\end{equation}
Exponential time-variables are conveniently defined by the following relation
\begin{equation}
z_m = z_m (t) = \frac{m}{H} e^{2Ht},
\hspace{0.5cm}
m \in \mathbb{N}^{*}.
\label{zm-var}
\end{equation} 
The expansion coefficients $\alpha^{i}_{m}$ and 
$\beta^{i}_{m}$ do not depend on time. As usual, by applying the canonical quantization method, the coefficients $\alpha^{i}_{m}$ and $\beta^{i}_{m}$ are promoted to operators. The canonically conjugate fields and the time-independent operators must satisfy simultaneously the equal time commutation relations and the oscillator commutators, respectively, that are given by the following relations
\begin{align}
\left[ x^{k} (t, \sigma) , \pi^{l}(t, \sigma') \right] & = i \delta^{kl} \delta(\sigma - \sigma' ),
\hspace{1.5cm}
\left[ x^{k}_{0} , \pi^{l}_{0} \right] = i \delta^{kl},
\label{ETC-fields}
\\
\left[
\alpha^{i}_{m}, \alpha^{j \dagger}_{n}
\right] & =
\left[
\beta^{i}_{m}, \beta^{j \dagger}_{n}
\right] = \delta^{ij} \delta_{mn},
\hspace{0.5cm}
\left[
\alpha^{i}_{m}, \beta^{j}_{n}
\right] = 0,
\label{ab-commut-rel}
\end{align}
for all $i, j =  1, 2, 3$ and all $m, n > 0$. These equations are satisfied in the cosmological gauge but not in an arbitrary time-dependent gauge. The above relations show that the two sets of operators $\alpha$ and $\beta$, respectively, should be interpreted as oscillating modes that move along the embedded circle $x(S^1)$ in opposed directions. These modes have time-dependent frequencies and can be paired according to the values of the indices $i$ and $m$ as in the flat space-time. The quantum dynamics is constrained to obey the following equations
\begin{align}
\mathit{H} (t)  & =  \frac{(p^{i}_0)^2}{4} +   h^{(2)}_{0} (t) + \sum_{m>0}
\left[
\Omega_{m} (z_m) 
\left(
\alpha^{i \dagger}_{m} \alpha^{i}_{m}
+
\beta^{i \dagger}_{m} \beta^{i}_{m}
\right)
+
\Phi^{(1)}_{m} (z_m)  \alpha^{i}_{m} \beta^{i}_{m}
+
\Phi^{(2)}_{m} (z_m)  \alpha^{i \dagger}_{m} \beta^{i \dagger}_{m} 
\right]=0, 
\label{quantum-constr-hamilt}
\\
\mathit{P} (t)  & =  \frac{\pi}{H}\sum_{m > 0}
m 
\frac{d J_{0}(z_m)}{d z_m} Y_{0}(z_m)
\left( 
\alpha^{i \dagger }_{m} \alpha^{i}_{m} 
-
\beta^{i \dagger }_{m} \beta^{i}_{m}
\right)=0,
\label{quantum-constr-level}
\end{align} 
where
\begin{equation}
h^{(2)}_{0} (t) = \frac{1}{4} e^{-4Ht} - e^{-2Ht} , 
\label{const-phi-h}
\end{equation}
is the contribution of the zero modes to $\mathit{H} (t)$. The costraints (\ref{quantum-constr-hamilt}) and (\ref{quantum-constr-level}) result from the  the vanishing of the energy-momentum tensor $T_{ab}=0$ \cite{Li:2007gf}. In the above relations, all products of oscillator operators are normal ordered. Their coefficients from the right hand side of the equation (\ref{quantum-constr-hamilt}) are given by the following time-dependent functions 
\begin{align}
\Omega_{m} (z_m) & =  \pi H z^{2}_{m} 
\left[
2 \left( \frac{d Y_{0}(z_m)}{d z_m}\right)^{2}
+ \frac{1}{2} \left( Y_{0}(z_m) \right)^{2}
\right],
\label{Omega}
\\
\Phi^{(1)}_{m} (z_m) & =  i\pi H z^{2}_{m} 
\left[
2  \frac{d Y_{0}(z_m)}{d z_m} \frac{d H^{(2)}_{0}(z_m)}{d z_m}
+ 
\frac{1}{2} Y_{0}(z_m) H^{(2)}_{0}(z_m)
\right]
\label{Phi-1}
\\
\Phi^{(2)}_{m} (z_m) & = - i\pi H z^{2}_{m} 
\left[
2  \frac{d Y_{0}(z_m)}{d z_m} \frac{d H^{(1)}_{0}(z_m)}{d z_m}
+ 
\frac{1}{2} Y_{0}(z_m) H^{(1)}_{0}(z_m),
\right]
\label{Phi-2}
\end{align}
where $a = 1, 2$ and $Y_{0}(z_m)$ is the Bessel function of the second kind. It was shown in \cite{Vancea:2016tkt} that the operator $\mathit{P} (t)$ can be interpreted as the generator of the translations along the spatial direction of the embedded cylinder. On the other hand, the operator $\mathit{H} (t)$ is the generator of the entanglement between the $\alpha$ and $\beta$ sectors, respectively, that occurs during their local time evolution and it is identical to the canonical Hamiltonian up to a smooth time-dependent function. The entanglement is caused by the interaction of the fields of the $\sigma$-model with the components of the metric on the $dS_4$ and it depends on time. The mapping of the entangled oscillator states between two instants of time can be described by the partial evolution map $\Upsilon (t_2,t_1)$ that is generated by the operator 
$\mathit{H} (t)$ and it is given by the following relation  
\begin{equation}
\Upsilon (t_2,t_1)
 =  \mathcal{T}
\left[ 
\exp\left(-i 
	\int^{t_2}_{t_1} dt  \mathit{H}(t)
	 \right)
\right]. 
\label{part-evol-map}
\end{equation}
The operator $\Upsilon (t_2,t_1)$ acts on the states at time $t_1$ and maps them to 
states at time $t_2$, respectively. The time-dependent states belong to the one-parameter family of time-dependent Hilbert spaces $\mathbf{H} (\mathbb{R}^{+})$ that has the structure of a direct product 
\begin{equation}
\mathbf{H} (\mathbb{R}^{+}) = \{ \mathcal{H}_t \}_{t \in \mathbb{R}^{+}}
= \mathcal{H}_{0} \bigotimes 
\left[
\mathsf{F} (\mathbb{R}^{+}) \bigotimes \mathcal{H}.
\right]
\label{Hilbert-space-family}
\end{equation}
The space $\mathcal{H}_{0} (\mathbb{R}^{+})$ contains the time-independent eigenstates of the operators  $ x^{i}_{0}$ and  $p^{i}_{0}$. The time-independent oscillator states belong to the Hilbert space $\mathcal{H}$. Their linear combinations are formed with coefficients from $\mathsf{F} (\mathbb{R}^{+})$ that denotes smooth time-dependent functions defined in the north pole diamond of the de Sitter space. At any given instant of time, a state of the system is represented by a superposition of elements from $\mathcal{H}$ with coefficients that belong to the set $\mathsf{F}(t) = \mathsf{F} (\mathbb{R}^{+})|_{t}$. Two operators at different values of time that belong to an one-parameter family of operators $\{ \mathcal{O}(t) \}_{t \in \mathbb{R}^{+}}$ are mapped into one another by the induced partial map
\begin{equation}
\mathcal{O}(t_1) \longmapsto \mathcal{O}(t_2) = 
\Upsilon (t_2,t_1) \mathcal{O}(t_1) 
\Upsilon (t_2,t_1)^{\dagger}.
\label{partial-map-induced}
\end{equation}
The equations (\ref{part-evol-map}) and (\ref{partial-map-induced}) determine the partial evolution of the $\sigma$-model in the north pole diamond of the de Sitter space as seen by the comoving observer\footnote{As discussed in \cite{Vancea:2016tkt}, this map does not describe the complete evolution of the system which must also take into account the evolution of the background metric and would ideally be derived from a theory of the Quantum Gravity.}. The physical states belong to the physical subspace $\mathbf{H}_{phys} (\mathbb{R}^{+})$ of $\mathbf{H} (\mathbb{R}^{+})$. This is defined by the following relation 
\begin{equation}
\mathbf{H}_{phys} (\mathbb{R}^{+}) = \mbox{ker}H[\mathbf{H} (\mathbb{R}^{+})]
\bigcap \mbox{ker}P[\mathbf{H} (\mathbb{R}^{+})].
\label{physical-H-subspace}
\end{equation}
For the explicit calculation of $\mathbf{H}_{phys} (\mathbb{R}^{+})$ we reefer to \cite{Vancea:2016tkt}.

\subsection{The untangled modes}

The normal ordered operator $\mathit{H}(t)$ from the equation (\ref{quantum-constr-hamilt}) contains non-diagonal terms that generate the entanglement between the $\alpha$ and $\beta$ sectors, respectively. Moreover, the time-independent eigenstates of the oscillators from these sectors are not eigenstates of $\mathit{H} (t)$. 

In order to find a basis in which the operator $\mathit{H} (t)$ is diagonal, we consider the general Bogoliubov transformations that act on the pair $(\alpha^{i}_{m} , \beta^{i}_{m})$ with definite values of indices $(i,m)$ according to the following equations
\begin{align}
a^{i}_{m} (z_m)& =  \cosh \theta_m (z_m) e^{i(\delta_m (z_m)+\varphi_m (z_m)) } \alpha^{i}_{m}  
                 +  \sinh \theta_m (z_m) e^{i\varphi (z_m)} \beta^{i\dagger}_{m},
\label{Bogoliubov-1}
\\
b^{i}_{m} (z_m)& =  \cosh \theta_m (z_m) e^{i(\delta_m (z_m) +\psi_m (z_m))}  \beta^{i}_{m}
                 +  \sinh \theta_m (z_m) e^{i\psi_m (z_m)} \alpha^{i\dagger}_{m} ,
\label{Bogoliubov-2}
\end{align}
where the parameters $\theta_m (z_m)$, $\delta_{m}(z_m)$, $\varphi_m (z_m)$ and $\psi_m (z_m)$ are real functions on time. They depend only on the oscillating mode $m$ and have the same form for all spatial directions. The operators $a^{i}_{m} (z_m)$ and $b^{i}_{m} (z_m)$ satisfy the standard commutation relations
\begin{equation}
[a^{i}_{m}(z_m) , a^{j\dagger}_{n} (z_n)] = [b^{i}_{m} (z_m), b^{j\dagger}_{n}(z_n) ] = \delta^{ij} \delta_{mn}.
\label{ETC-a-b}
\end{equation}
The operator $\mathit{H} (t)$ can be expressed in the $a-b$ representation by applying the transformations 
(\ref{Bogoliubov-1}) and (\ref{Bogoliubov-2}) to the right hand side of the equation (\ref{quantum-constr-hamilt}). 
After some algebraic manipulations, we arrive at the following diagonal terms
\begin{equation}
\mathit{H} (t) = \frac{(p^{i}_0)^2}{4} + h^{(2)}_{0} (t) + \sum_{m>0}
f_{m} (z_m) 
\left[
a^{i \dagger}_{m} (z_m) a^{i}_{m} (z_m)
+
b^{i \dagger}_{m} (z_m) b^{i}_{m} (z_m)
\right],
\label{H-operator-diagonal}
\end{equation}
where the frequencies $f_{m} (z_m)$ of the $a-b$ modes are given by the following real functions on time
\begin{equation}
f_m (z_m) = \Omega_{m} (z_m)\cosh (2 \theta_m (z_m)) 
- \mathfrak{Re}\left[\Phi^{(1)}_m(z_m ) e^{i\delta_m (z_m)}\right]\sinh (2\theta_m (z_m)).
\label{frequencies-untangled}
\end{equation}
The frequencies $f_m (z_m)$ are positive for all values of $\theta_m(z)$ if $\cos(\lambda_m(z_m)) \geq 0$. If $\cos(\lambda_m(z_m)) \geq 0 < 0$, the oscillator frequencies are negative. The requirement that the non-diagonal part of $\mathit{H} (t)$ vanish in the $a-b$ representation imposes the following constraints on the parameters of the Bogoliubov transformations
\begin{align}
\tanh \theta_{m} & = 
\frac{\cos \lambda_m  \cos \delta_m  \pm 
\sqrt{ \cos^2 \lambda_m  \cos^2 \delta_m  - \cos \lambda_m  \cos (\lambda_m -2 \delta_m )} }{\cos (\lambda_m )}
\label{theta-constraint}
\\
& 
=
\frac{\cos \lambda_m  \sin \delta_m  \mp 
\sqrt{ \cos^2 \lambda_m  \sin^2 \delta_m  + \sin \lambda_m  \sin (\lambda_m -2 \delta_m )} }{\sin \lambda_m }
,
\label{reality-constraint}
\end{align}
where we have introduced the notation $\lambda_m = \arg(\Phi^{(1)}_{m}(z_m))$ and we have dropped the variable $z_m$ for simplicity. The first equation above simply expresses the parameter $\theta_{m}(z_m)$ in terms of other parameters of the Bogoliubov transformations and the  Hankel functions. The parameters $\varphi_m (z_m)$ and $\psi_m (z_m)$ are left arbitrary by these conditions. The fact that the parameters of the Bogoliubov transformations are functions on time implies that the modes $a^{i}_{m}(z_m)$ and $b^{i}_{m}(z_m)$ depend on time, too. Therefore, the commutation relations (\ref{ETC-a-b}) should be understood at equal times. The second constraint (\ref{reality-constraint}) is necessary to guarantee that the functions $\tanh \theta_{m} (z_m ) $ are real and it determines the parameters $\delta_m(z_m )$'s as functions of $\lambda_m(z_m )$'s which can be calculated from the relation (\ref{Phi-1}). After some manipulations of the nonlinear algebraic equation (\ref{reality-constraint}), one can show that the trigonometric functions of $\delta_m(z_m )$ are real and of modulus smaller than one. This proves that the constraints are consistent with the properties of their variables.

The Fock space of the $\sigma$-model at any given value of time is spanned by linear combinations of multi-excitation states of the form
\begin{equation}
| \Psi (t) \rangle_{osc} = \sum_{\mathcal{N}_{a}(t)} \sum_{\mathcal{N}_{b}(t)}
C(\mathcal{N}_{a}(t) ; \mathcal{N}_{b}(t)) | \mathcal{N}_{a}(t) ; \mathcal{N}_{b}(t) \rangle ,
\label{gen-vect}
\end{equation}
where we have used the standard multi-index notation
\begin{align}
\mathcal{N}_{a}(t) & =  
\{ N^{i}_{a, m} (z_m) \}^{i = 1, 2, 3}_{m > 0} =
\{ N^{1}_{a, 1} (z_m) , N^{2}_{a, 1} (z_m), N^{3}_{a, 1} (z_m);
N^{1}_{a, 2} (z_m), N^{2}_{a, 2} (z_m), N^{3}_{a, 2}(z_m);
\cdots
\} ,
\label{multi-index-t-a}
\\
\mathcal{N}_{b}(t) & =  
\{ N^{i}_{b, m} (z_m) \}^{i = 1, 2, 3}_{m > 0} =
\{ N^{1}_{b, 1}(z_m), N^{2}_{b, 1}(z_m), N^{3}_{b, 1}(z_m);
N^{1}_{b, 2} (z_m), N^{2}_{b, 2}(z_m) , N^{3}_{b, 2}(z_m);
\cdots
\} .
\label{multi-index-t-b}
\end{align}
Here, $N^{i}_{a, m}(z_m)$ and $N^{i}_{b, m}(z_m)$ denote the eigenvalues of the corresponding number operators $\hat{N}^{i}_{a, m}(z_m)$ and $ \hat{N}^{i}_{b, m}(z_m)$, respectively. The coefficients $C(\mathcal{N}_{a}(t) ; \mathcal{N}_{b}(t))$ are smooth complex functions on the local time parameter. The vectors from the Fock space are obtained in the usual fashion by applying repeatedly the creation operators on the ground state of the oscillator system $| 0 (t) \rangle = \prod |0 (z_m) \rangle $ defined by the relations
\begin{equation}
a^{i \dagger}_{m} (z_m(t)) |0 (t) \rangle = b^{i\dagger}_{m} (z_m) |0 (t) \rangle = 0, 
\label{vac-t}
\end{equation}
for all $m>0$ and all $i= 1, 2, 3$. As expected, the Fock space at a given time has the structure of the direct product
\begin{equation}
\mathfrak{F}_{} (t) = \bigotimes_{i,m} \mathfrak{F}_{(i,m)} (z_m).
\label{Fock-t}
\end{equation} 
Thus, the full Hilbert space of the $\sigma$-model in the untangled representation $a-b$ is a one-parameter family of Hilbert spaces  
\begin{equation}
\mathfrak{H} (\mathbb{R}^{+}) = \{ \mathfrak{H} (t) \}_{t \in \mathbb{R}^{+}} =
\mathfrak{H}_{0} \bigotimes \{ \mathfrak{F}_{} (t) \}_{t \in \mathbb{R}^{+}},
\label{Hilbert-space-fin}
\end{equation}
where $\mathfrak{H}_{0}$ is the Hilbert space of the zero mode operators $x^{i}_{0}$ and $p^{i}_{0}$, respectively. The physical subspace $\mathfrak{H} (\mathbb{R}^{+})_{phys} \in \mathfrak{H} (\mathbb{R}^{+})$ is defined as 
\begin{equation}
\mathfrak{H}_{phys} (\mathbb{R}^{+}) = \mbox{ker}\mathit{H}[\mathfrak{H} (\mathbb{R}^{+})]
\bigcap \mbox{ker}\mathit{P}[\mathfrak{H} (\mathbb{R}^{+})].
\label{physical-H-family}
\end{equation}
By applying the Bogoliubov transformations (\ref{Bogoliubov-1}) and (\ref{Bogoliubov-2}) to the right hand side of the equation (\ref{quantum-constr-level}), one can see that the operator $\mathit{P} (t)$ takes the following form
\begin{equation}
\mathit{P} (t)  =  \frac{\pi}{H}\sum_{m > 0}
m 
\frac{d J_{0}(z_m)}{d z_m} Y_{0}(z_m)
\left[
a^{i \dagger }_{m}(z_m)  a^{i}_{m} (z_m)
-
b^{i \dagger }_{m} (z_m) b^{i}_{m} (z_m)
\right].
\label{P-a-b}
\end{equation}
This shows that the space $\mathfrak{H}_{phys} (\mathbb{R}^{+})$ contains just states that satisfy the level matching condition at every value of time. On the other hand, there is a relationship among eigenvalues of different operators calculated in states that belong to the kernel $\mbox{ker}\mathit{H}[\mathfrak{H} (\mathfrak{R}^{+})]$, namely,  
\begin{equation}
\mathcal{E} (t) - \frac{(p^{i}_0)^2}{4} = 3\sum_{m>0}
f_{m} (z_m) 
\left[
N^{i}_{a,m}(z_m) + N^{i}_{b,m}(z_m)
\right]
+  h^{(2)}_{0} (t),
\label{E-P-t}
\end{equation}
where $\mathcal{E} (t)$ is the eigenvalue of $\mathit{H}(t)$. The equation (\ref{E-P-t}) is an energy-momentum-like relation of which outcomes could be observed locally in the co-moving frame in the north pole of the de Sitter space in the cosmological gauge. Note that the last term in the right hand side of the above equation is the result of the interaction between the oscillators and the de Sitter background metric. In the case of the ground state, the equation (\ref{E-P-t}) reduces to
\begin{equation}
\mathcal{E}_0 (t) - \frac{(p^{i}_0)^2}{4} = 
  h^{(2)}_{0} (t).
\label{E-P-t-0}
\end{equation}
The lowest excited oscillator state corresponds to $m=1$ and has the following form 
\begin{equation}
C_{ij}(t) a^{i \dagger}_{1} (z_1(t)) 
b^{j \dagger}_{1} (z_1(t)) |0 (t) \rangle.
\label{lowest-state}
\end{equation}
For these states, the equation (\ref{E-P-t}) takes the following form
\begin{equation}
\mathcal{E}_{1} (t) - \frac{(p^{i}_0)^2}{4} = 6
f_{1} (z_1(t)) 
+  h^{(2)}_{0} (t).
\label{E-P-t-1}
\end{equation}
Since the quantities from the right hand side of the equations (\ref{E-P-t-0}) and (\ref{E-P-t-1}) are fixed, these equations impose restrictions on the values of $\mathcal{E} (t)$ and  $ p^{i}_0$. Similar relations can be derived for states with a higher number of excitations.

We end this section with a brief comment on the time-variation of the untangled oscillators. In this representation, the partial evolution operator from the equation (\ref{part-evol-map}) is written in terms of the diagonal operator 
$\mathit{H}(t)$ from the relation (\ref{H-operator-diagonal}). It is easy to see that the action of the operator $\Upsilon (t_2,t_1)$ on the operators $a^{i}_{m}(z_m (t))$ and $b^{i}_{m}(z_m (t))$ given by the equation (\ref{partial-map-induced}) is compatible with the following Heisenberg-like equations
\begin{equation}
\frac{d a^{i}_{m}(z_m (t))}{dt} = 
i \left[ 
a^{i}_{m} (z_m (t)), \mathit{H} (t) 
\right],
\hspace{0,5cm} 
\frac{d b^{i}_{m}(z_m (t))}{dt} = 
i \left[ 
b^{i}_{m} (z_m (t)), \mathit{H} (t) 
\right],
\label{Heisenberg-equations}
\end{equation}
for all $i = 1, 2, 3$ and all $m>0$. The above equations together with the partial map of states 
\begin{equation}
| \Psi (t_2) \rangle = \Upsilon (t_2 , t_1) | \Psi (t_1) \rangle
\label{states-equation}
\end{equation}
determine the partial evolution of the untangled excitations of the $\sigma$-model.

\section{Nonequilibrium dynamics of the untangled modes}

In the previous section, we have shown that the left and right moving modes on the embedded base space $x(\Sigma^{1,1})$ can be untangled from each other. However, each of the untangled modes still interacts with the components of the de Sitter metric. The most important consequence of this fact is that the number of quanta of a given mode can vary with time as observed in the local frame associated with the embedded cylinder $x(\Sigma^{1,1})$. Moreover, the frequencies of all modes $f_m (z_m (t))$ are time-dependent. This shows that the untangled modes are subject to nonequilibrium dynamical equations that describe their evolution under the continuous exchange of energy with the classical time-dependent metric fields. In this section, we are going to determine these equations. Since  the problem is formulated in the canonical formalism, we will apply the NETFD method that can be found in \cite{Umezawa:1993yq,Mizutani:2011rt,Mizutani:2011jy,Nardi:2011sk}.

\subsection{The thermal doublet modes}

We begin by constructing the thermal doublets corresponding to the untangled modes of the $\sigma$-model. To this end, we associate to each oscillator an identical copy of it, denoted by a tilde. Then the Hilbert space $\hat{\mathfrak{H}} (\mathbb{R}^{+})$ of the total system is the direct sum of the Hilbert spaces of all untangled oscillators and their copies
\begin{equation}
\hat{\mathfrak{H}} (\mathbb{R}^{+}) = \mathfrak{H} (\mathbb{R}^{+}) 
\otimes \tilde{\mathfrak{H}} (\mathbb{R}^{+}).
\label{total-Hilbert-space}
\end{equation} 
The above relation shows that for every pair of indices $(i,m)$ there are two pairs of independent oscillators 
$(a^{i}_{m}(z_m), \tilde{a}^{i}_{m}(z_m))$ and $(b^{i}_{m}(z_m), \tilde{b}^{i}_{m}(z_m))$, respectively, that act on the corresponding total Fock space which can be decomposed as
\begin{align}
\hat{\mathfrak{F}}_{(i,m)}(z_m) & = \mathfrak{F}_{(i,m)}(z_m) \otimes \tilde{\mathfrak{F}}_{(i,m)}(z_m) 
\nonumber
\\
                                & = \left[ 
                                     \mathfrak{F}^{a}_{(i,m)}(z_m)                                    
                                    \otimes
                                    \mathfrak{F}^{b}_{(i,m)}(z_m)                                    
                                    \right]
                                    \otimes
                                     \left[ 
                                     \tilde{\mathfrak{F}}^{a}_{(i,m)}(z_m)                                    
                                    \otimes
                                    \tilde{\mathfrak{F}}^{b}_{(i,m)}(z_m)                                    
                                    \right].
\label{therm-Fock-space}
\end{align}
The oscillator operators satisfy the standard canonical commutation relations 
\begin{align}
[a^{i}_{m}(z_m) , a^{j\dagger}_{n} (z_n)] & = [\tilde{a}^{i}_{m} (z_m), \tilde{a}^{j\dagger}_{n}(z_n) ] = \delta^{ij} \delta_{mn} ,
\label{ETC-double-aa}
\\
[b^{i}_{m}(z_m) , b^{j\dagger}_{n} (z_n)] & = [\tilde{b}^{i}_{m} (z_m), \tilde{b}^{j\dagger}_{n}(z_n) ] = \delta^{ij} \delta_{mn} ,
\label{ETC-double-bb}
\\
[a^{i}_{m}(z_m) , \tilde{a}^{j}_{n} (z_n)] & = [a^{i}_{m} (z_m), \tilde{b}^{j}_{n}(z_n) ] = \ldots = 0.
\label{ETC-double-ab}
\end{align}
Since the oscillators are time-dependent, the above relations should be understood as holding at equal times. The states from $\hat{\mathfrak{F}}_{(i,m)} (z_m)$ are constructed as usual by acting with the creation operators on the total vacuum 
\begin{equation}
| 0 (z_m ) \rangle \rangle = | 0 (z_m ) \rangle_{a} \otimes | 0 (z_m ) \rangle_{b}
\label{therm-vacuum}
\end{equation}
which is annihilated by all annihilation operators from the tilde as well as the non-tilde sectors, respectively,
\begin{equation}
a^{i}_{m}(z_m)| 0 (z_m ) \rangle \rangle = \tilde{a}^{i}_{m}(z_m) | 0 (z_m ) \rangle \rangle =
b^{i}_{m}(z_m) | 0 (z_m ) \rangle \rangle =  \tilde{b}^{i}_{m}(z_m)| 0 (z_m ) \rangle \rangle = 0,
\label{annih-therm-vac}
\end{equation}
for all $m > 0$ and $i = 1, 2, 3$. In what follows, we are going to use the notation in which the time variable $t$ is explicit instead of $z_m$. We recall that the relationship between the two variables is given by the equation (\ref{zm-var}) above. 

The time-dependent thermal doublets are two dimensional vectors with operator entries defined by the following relations
\begin{align}
a^{i \alpha}_{m} (t) & =
\begin{pmatrix}
a^{i}_{m} (t)
\\
\tilde{a}^{i\dagger}_{m} (t)
\end{pmatrix}
\, ,
\hspace{0.5cm}
\bar{a}^{i \alpha}_{m} (t) =
\left(
a^{i\dagger}_{m} (t)
\hspace{0.2cm}
-\tilde{a}^{i}_{m} (t)
\right),
\label{a-double}
\\
b^{i \alpha}_{m} (t) & =
\begin{pmatrix}
b^{i}_{m} (t)
\\
\tilde{b}^{i\dagger}_{m} (t)
\end{pmatrix}
\, ,
\hspace{0.5cm}
\bar{b}^{i \alpha}_{m} (t) =
\left(
b^{i\dagger}_{m} (t)
\hspace{0.2cm}
-\tilde{b}^{i}_{m} (t)
\right),
\label{b-double}
\end{align}
where $\alpha, \beta = 1, 2$ are two-dimensional vector indices. The row vectors can be obtained from the column vectors if a bar-conjugation operation 
\begin{equation}
\bar{a}^{\alpha} = (a^{\dagger})^{\beta} s^{\beta \alpha}_{3},
\label{bar-op}
\end{equation}
where $s_3$ stands for the corresponding Pauli matrix. The action of the thermal doublets on the vacuum is defined by the action of their components as given by the equation (\ref{annih-therm-vac}) above. 

The NETFD prescription \cite{Umezawa:1993yq} requires the introduction of a time-independent pseudo-particle doublet for each thermal doublet that is defined by the following relations
\begin{align}
\xi^{i\alpha}_{m} & = \exp \left[ i \int^{t}_{t_{0}} ds \, q_m (s)  \right]
B^{\alpha \beta}_{a,m} (t) a^{i\beta}_{m} (t),
\label{xi-doubles-1}
\\
\bar{\xi}^{i\alpha}_{m} & = \bar{a}^{i\beta}_{m} (t)
\exp  \left[ - i \int^{t}_{t_{0}} ds \, q_m (s) \right]
(B^{-1}_{a,m})^{\beta \alpha} (t),
\label{xi-doubles-2}
\\
\chi^{i\alpha}_{m} & = \exp \left[ i \int^{t}_{t_{0}} ds \, q_m (s)  \right]
B^{\alpha \beta}_{b,m} (t) b^{i\beta}_{m} (t),
\label{chi-doubles-1}
\\
\bar{\chi}^{i\alpha}_{m} & = \bar{b}^{i\beta}_{m} (t)
\exp  \left[ - i \int^{t}_{t_{0}} ds \, q_m (s) \right]
(B^{-1}_{b,m})^{\beta \alpha} (t),
\label{chi-doubles-2}
\end{align}
where the Bogoliubov matrices are given by the following relations (see the Appendix)
\begin{align}
B_{a,m} (t) & = \left( \begin{array}{cc}
1 + n_{a,m} (t) & -  n_{a,m} (t) \\
-1            & 1            \\
\end{array} 
\right),
\label{Bogolibov-matrix-a}
\\
B_{b,m} (t) & = \left( \begin{array}{cc}
1 + n_{b,m} (t) & -  n_{b,m} (t) \\
-1            & 1            \\
\end{array} 
\right).
\label{Bogolibov-matrix-b}
\end{align}
The time-independent operators are necessary to define the time-independent pseudo-particle vacuum $|0 \rangle $ in which all physical quantities should be computed \cite{Umezawa:1993yq}. For example, the number of untangled excitations from the equations (\ref{Bogolibov-matrix-a}) and (\ref{Bogolibov-matrix-b}), respectively, are defined by the vacuum expectation value of the thermal doublet operators in the pseudo-particle vacuum 
\begin{align}
n_{a,m} (t) \delta^{ij} \delta_{ml} & = 
\langle 0 | \bar{a}^{i 1}_{m}(t) a^{j 1}_{l} (t) | 0 \rangle ,
\label{number-average-a}
\\
n_{b,m} (t) \delta^{ij} \delta_{ml} & = 
\langle 0 | \bar{b}^{i 1}_{m}(t) b^{j 1}_{l} (t) | 0 \rangle .
\label{number-average-b}
\end{align} 
In order to determine the evolution of the thermal doublets under the action of the partial evolution map $\Upsilon (t_2,t_1)$ one needs to construct a second set of time-dependent pseudo-particle operators 
 $\xi^{i\alpha}_{m} (t)$ and $\chi^{i\alpha}_{m} (t)$ that differ from $\xi^{i\alpha}_{m}$ and $\chi^{i\alpha}_{m}$, respectively, by the time-dependent phases
\begin{align}
\xi^{i\alpha}_{m} (t) & = \exp \left[ i \int^{t}_{t_{0}} ds \, r_m (s)  \right]
\xi^{i\alpha}_{m} ,
\label{xi-time-1}
\\
\bar{\xi}^{i\alpha}_{m} (t) & = \bar{\xi}^{i\alpha}_{m} (t)
\exp  \left[ - i \int^{t}_{t_{0}} ds \, r_m (s) \right],
\label{xi-time-2}
\\
\chi^{i\alpha}_{m} (t) & = \exp \left[ i \int^{t}_{t_{0}} ds \, r_m (s)  \right]
\chi^{i\alpha}_{m} ,
\label{chi-time-1}
\\
\bar{\chi}^{i\alpha}_{m} (t) & = \bar{\chi}^{i\alpha}_{m} (t)
\exp  \left[ - i \int^{t}_{t_{0}} ds \, r_m (s) \right].
\label{chi-time-2}
\end{align}
Two pseudo-particle doublets $\xi^{i\alpha}_{m} (t)$ and $\xi^{i\alpha}_{m}$, respectively, are associated to the same untangled doublet $a^{i\alpha}_{m}(t)$ if $q_m (s) = r_m (s)$. In this case, the Bogoliubov matrix defined by the equation (\ref{Bogolibov-matrix-a}) maps $a^{i\alpha}_{m}(t)$ to  $\xi^{i\alpha}_{m} (t)$ as can be easily verified. 
After some algebra, one can show that the operators  $a^{i\alpha}_{m}(t)$ and  $\bar{a}^{i\alpha}_{m}(t)$ satisfy the following partial evolution equations
\begin{align}
\left[
 \delta^{\alpha \beta} \left( \frac{d}{dt} - i f_{m} (t) \right) 
+ Q^{\alpha \beta}_{a,m} (t)
\right] a^{i \beta}_{m} (t) & = 0,
\label{a-evol}
\\
\bar{a}^{i \beta}_{m} (t)
\left[
 \delta^{\beta \alpha} \left( \frac{d}{dt} + i f_{m} (t) \right) 
- Q^{\beta \alpha}_{a,m} (t)
\right]  & = 0,
\label{a-bar-evol}
\end{align}
where we have introduced the following notation
\begin{equation}
Q^{\alpha \beta}_{a,m} (t) = (B^{-1}_{a,m})^{\alpha \gamma}(t)  \frac{d B_{a,m}^{\gamma \beta} (t) }{dt}.
\label{P-op}
\end{equation}
The equations that describe the partial evolution of the operators  $b^{i\alpha}_{m}(t)$ and  $\bar{b}^{i\alpha}_{m}(t)$ can be obtained from the equations (\ref{a-evol}) and (\ref{a-bar-evol}), respectively, by replacing to quantities that reefer to the $a$-modes by the ones corresponding to the $b$-modes. The equations (\ref{a-evol}) and (\ref{a-bar-evol}) should be compared with the action of the partial evolution map from the equation (\ref{part-evol-map}) that characterizes the dynamics of the $\sigma$-model in the de Sitter background \cite{Vancea:2016tkt}. By using the equation (\ref{partial-map-induced}), one can easily see that the normal ordered operator $\mathit{H}_{a} (t)$, generates the first two terms from each of the equations (\ref{a-evol}) and (\ref{a-bar-evol}) if $q_m (t) = f_m (t)$. These terms define the evolution of the pseudo-particle operators $\xi^{i\alpha}_{m} (t)$ and $\bar{\xi}^{i\alpha}_{m} (t)$, respectively. The last term is determined by the operator $Q^{\alpha \beta}_{a,m} (t)$ that contains the contribution of the Bogoliubov matrix to $\mathit{H}_{a} (t)$ and which is also present in the interaction theory where it generates a counterterm in the interacting Hamiltonian \cite{Umezawa:1993yq}. The same considerations can be made for the excitations from the $b$-sector and we reach at similar conclusions.

This analysis shows that the nonequilibrium dynamics of the untangled modes has the standard NETFD form for time-dependent system of oscillators. The dynamics is determined in terms of the operators $\mathit{H}_{a(b)} (t)$ and $\sum_{m} Q_{a(b),m} (t)$ that generate together the following nonequilibrium evolution map of the total system
\begin{equation}
\hat{\mathit{H}}_{tot,osc}(t) = \hat{\mathit{H}}(t) + \hat{\mathit{Q}} (t),
\label{full-evolution-operator}
\end{equation} 
where
\begin{align}
\hat{\mathit{H}} (t) & =  \sum_{m>0}
f_{m} (t) \, \delta_{\alpha \beta} \,
\left[
\bar{a}^{i\alpha}_{m} (t) \, a^{i\beta}_{m} (t) +
\bar{b}^{i\alpha}_{m} (t) \, b^{i\beta}_{m} (t)
\right],
\\
\hat{\mathit{Q}} (t) & =  \sum_{m>0}
\left[
Q^{\alpha \beta}_{a,m} (t) \,
\bar{a}^{i\alpha}_{m} (t) \, a^{i\beta}_{m} (t) +
Q^{\alpha \beta}_{b,m} (t) \,
\bar{b}^{i\alpha}_{m} (t) \, b^{i\beta}_{m} (t) 
\right].
\end{align}
Here, the operator $Q^{\alpha \beta}_{b,m} (t)$ has the same form as the operator $Q^{\alpha \beta}_{a,m} (t)$ but with 
$B_{b,m}(t)$ replacing $B_{a,m}(t)$ in the equation (\ref{P-op}) above.

By extending in a simple manner the NETFD formalism, one can associated thermal doublets to the degrees of freedom $x^{i}_{0}$ and $p^{i}_{0}$ that are necessary to describe a full copy of the original $\sigma$-model \cite{Vancea:2016tkt}. To this end, we firstly define the following variables
\begin{equation}
X^{i}_{0} (t) = x^{i}_{0} + p^{i}_{0} t,
\hspace{0.5cm}
\Pi^{i}_{0} = \frac{1}{2\pi} p^{i}_{0}.
\label{zero-modes}
\end{equation}
Next, we introduce the zero mode oscillator operators $a^{i}_{0}$ and $a^{i \dagger}_{0}$ that satisfy the canonical commutation relations $[a^{i}_{0},a^{j \dagger}_{0}]= \delta^{ij}$. Then it is easy to show that the equations
(\ref{zero-modes}) can be rewritten as follows
\begin{align}
X^{i}_{0} (t) & = \frac{1}{2}(\pi t + i) a^{i}_{0} + \frac{1}{2}(\pi t - i ) a^{i \dagger}_{0} ,
\label{x-zero-mode}
\\
\Pi^{i}_{0} & = \frac{1}{2 \pi} \left( a^{i}_{0} + a^{i \dagger}_{0}\right).
\label{p-zero-mode}
\end{align}
One can verify that the operators $X^{i}_{0} (t)$ and $\Pi^{i}_{0}$ obey the same canonical commutation relations in both representations. In order to construct the thermal degrees of freedom, we associate identical copies to the zero modes oscillators denoted by $\tilde{X}^{i}_{0} (t)$ and $\tilde{\Pi}^{i}_{0}$. These operators can be obtained from (\ref{x-zero-mode}) and (\ref{p-zero-mode}) by applying the tilde-conjugation rules from the Appendix. Then the thermal doublet of the zero modes is given by the equations (\ref{a-double}). The result can be written as follows  
\begin{align}
\phi^{i \alpha}_{0} (t) & = U(t) a^{i \alpha}_{0} + U^{*} (t) (s_3 \bar{a}^{i T}_{0})^{\alpha},
\label{zero-mode-double}
\\
\frac{d \phi^{i \alpha}_{0}}{dt} & = \frac{\pi}{2} \left( a^{i \alpha}_{0} + (s_3 \bar{a}^{i T}_{0})^{\alpha} \right).
\label{eq-mot-zero-mode-double}
\end{align}
where  $U(t)= (\pi t + i)/2$. The equations (\ref{a-evol}), (\ref{a-bar-evol}) and (\ref{eq-mot-zero-mode-double}) determine the partial evolution of the thermal doublets of all untangled modes of the $\sigma$-model including the center of the circle $x(S^1)$. They can be used to calculate physical quantities associated to local observations in a neighbourhood of the embedded base space as we are going to see in the next subsection.

\subsection{Free one-body propagator}

One important quantity to be calculated in nonequilibrium is the free one-body propagator of an oscillating mode $m$ between two different values of time $t_1$ and $t_2$. Since the $a$ and $b$-sectors do not interact with each other, there is just one propagator for each of these sectors and no cross-sector propagation. These are defined by the following vacuum expectation values 
\begin{align}
\Delta^{i\alpha,j\beta}_{a,m} (t_1 , t_2 ) \delta_{mn} 
& =
-i \langle 0 | 
T \left[
a_{m}^{i\alpha} (t_1) \bar{a}_{n}^{j\beta} (t_2 )
\right] 
| 0 \rangle
,
\label{propagator-def-a}
\\
\Delta^{i\alpha,j\beta}_{b,m} (t_1 , t_2 ) \delta_{mn} 
& =
-i \langle 0 | 
T \left[
b_{m}^{i\alpha} (t_1) \bar{b}_{n}^{j\beta} (t_2 )
\right] 
| 0 \rangle ,
\label{propagator-def-b}
\end{align}
where $| 0 \rangle $ is the vacuum state of the time-independent pseudo-particle excitations. The propagators $\Delta^{i\alpha,j\beta}_{a,m} (t_1 , t_2 ) \delta_{mn}$ can be calculated by using the relationship between the two sets of operators $(a_{m}^{\alpha} (t_1) ,\bar{a}_{n}^{\nu} (t_2 ))$ and $(\xi^{i\alpha}_{m}, \bar{\xi}^{i\alpha}_{m})$, respectively, as given by the equations (\ref{xi-doubles-1}) and (\ref{xi-doubles-2}) above. A short computation produces the following results 
\begin{align}
\Delta^{i\alpha,j\beta}_{a,m} (t_1 , t_2 ) & = 
\delta^{ij}
\left[B^{-1}_{a,m} (t_1) \Delta_{m} (t_1 , t_2 ) B_{a,m} (t_2) \right]^{\alpha \beta}, 
\label{free-one-body-prop-a}
\\
\Delta^{i\alpha,j\beta}_{b,m} (t_1 , t_2 ) & = 
\delta^{ij}
\left[B^{-1}_{b,m} (t_1) \Delta_{m} (t_1 , t_2 ) B_{b,m} (t_2) \right]^{\alpha \beta}, 
\label{free-one-body-prop-b}
\end{align}
where 
\begin{align}
\Delta_{m} (t_1 , t_2 ) =   
i 
\left(
\begin{smallmatrix} 
- \theta(t_1 - t_2) & 0
\\ 
0& \theta(t_2 - t_1) 
\end{smallmatrix}
\right)
e^{i \int_{t_1}^{t_2} dt 
f_m (\frac{m}{H} e^{2Ht} )
}.
\label{free-one-body-prop-c}
\end{align}
The explicit form of the components of the propagator can be obtained after a simple matrix algebra. The results in the $a$-sector take the following form
\begin{align}
\Delta^{i1,j1}_{a,m} (t_1 , t_2 ) & = 
 - i \delta^{ij} e^{ i \int_{t_1}^{t_2} f_m (\frac{m}{H} e^{2Ht}) dt}
\left[
\theta (t_1 - t_2 ) + n_{a,m} (t_1) \theta (t_2 - t_1 )
+ n_{a,m} (t_2) \theta (t_1 - t_2 ) 
\right] ,
\nonumber
\\
\Delta^{i1,j2}_{a,m} (t_1 , t_2 )  & = 
 i \delta^{ij} e^{ i \int_{t_1}^{t_2} f_m (\frac{m}{H} e^{2Ht}) dt}
 \left[
n_{a,m} (t_1) \theta (t_2 - t_1 )
+ n_{a,m} (t_2) \theta (t_1 - t_2 ) 
\right] ,
\nonumber
\\
\Delta^{i2,j1}_{a,m} (t_1 , t_2 ) & = 
- i \delta^{ij} e^{ i \int_{t_1}^{t_2} f_m (\frac{m}{H} e^{2Ht}) dt}
 \left[
\theta (t_1 - t_2 ) \left(1 + n_{a,m} (t_2) \right)
+
\theta (t_2 - t_1 ) \left(1 + n_{a,m} (t_1) \right)
\right] ,
\nonumber
\\ 
\Delta^{i2,j2}_{a,m} (t_1 , t_2 )  & = 
 i \delta^{ij} e^{ i \int_{t_1}^{t_2} f_m (\frac{m}{H} e^{2Ht}) dt}
 \left[
\theta (t_2 - t_1 ) + n_{a,m} (t_1) \theta (t_2 - t_1 )
+ n_{a,m} (t_2) \theta (t_1 - t_2 ) 
\right].
\label{one-body-prop-comp}
\end{align}
As before, the calculations and the results in the $b$-sector duplicate the ones in the $a$-sector. 
The above equations show that the propagator of an untangled mode between $t_1$ and $t_2$ is completely determined by its frequency $f_m(t)$ and the number of excitations $n_{a(b),m}(t)$. The partial evolution equation of the latter can be obtained from the equations (\ref{Heisenberg-equations}) and (\ref{number-average-a}). It is easy to verify that the results is 
\begin{equation}
\frac{d n_{a(b),m}(t)}{d t} = 2i f_{m} (t) \, n_{a(b),m} (t),
\label{evol-number}
\end{equation} 
where the frequencies of the untangled modes $f_{m}(t)$ are given by the equation (\ref{frequencies-untangled}).
If interactions among the modes are introduced in either $a$ or $b$ sectors, respectively, the propagators $\Delta^{i\alpha,j\beta}_{a,m} (t_1 , t_2 )$ and $\Delta^{i\alpha,j\beta}_{b,m} (t_1 , t_2 )$ receive corrections as the interactions change the number of excitations being created at each instant of time \cite{Umezawa:1993yq}. 

\section{The thermal doublet fields}

We now turn to the question whether the nonequilibrium dynamics of the untangled excitations can be described in terms of fields that satisfy the equations of motion of the $\sigma$-model. These fields should have a thermal doublet structure since their modes should be the thermal doublet oscillators constructed in the previous section. 

In order to construct the thermal fields of the $\sigma$-model, we start by expressing the solutions given by the equations (\ref{sol-x-herm}) and (\ref{sol-p-herm}), respectively, in the $a-b$ representation. The result takes the following form
\begin{align}
x^{i}\left( t,\sigma \right) & = x_{0}^{i}+p_{0}^{i}t
+\frac{i}{2}
\sqrt{\frac{\pi}{H}}
\sum _{m>0}\left[ a_{m}^{i}\left( z_{m}\right)e^{im\sigma } F_m \left( z_{m}\right) + 
a_{m}^{i\dagger}\left( z_{m}\right) e^{-im\sigma } F^{*}_{m} \left( z_{m}\right)\right]
\nonumber
\\
& -\dfrac {i} {2}\sqrt {\dfrac {\pi } {H}}
\sum _{m>0}\left[ b_{m}^{i}\left( z_{m}\right)e^{-im\sigma } G_m \left( z_{m}\right) + 
b_{m}^{i\dagger}\left( z_{m}\right) e^{im\sigma } G^{*}_{m} \left( z_{m}\right)\right]
,
\label{sol-x-ab}
\\
\pi^{i}\left( t,\sigma \right) & = \frac{1}{2\pi} p_{0}^{i}
+\frac{i}{2}
\sqrt{\frac{H}{\pi}}
\sum _{m>0}\left[ a_{m}^{i}\left( z_{m}\right)e^{im\sigma } \bar{F}_m \left( z_{m}\right) + 
a_{m}^{i\dagger}\left( z_{m}\right) e^{-im\sigma } \bar{F}^{*}_{m} \left( z_{m}\right)\right]
\nonumber
\\
& -\dfrac {i} {2}\sqrt {\dfrac {H } {\pi}}
\sum _{m>0}\left[ b_{m}^{i}\left( z_{m}\right)e^{-im\sigma } \bar{G}_m \left( z_{m}\right) + 
b_{m}^{i\dagger}\left( z_{m}\right) e^{im\sigma } \bar{G}^{*}_{m} \left( z_{m}\right)\right]
,
\label{sol-p-ab}
\end{align}
where we have introduced the following notations
\begin{align}
F_m \left( z_{m}\right) & = 
\cosh \theta_m (z_m) e^{-i(\delta_m (z_m)+\varphi_m (z_m))} H_{0}^{(2)}\left( z_{m}\right)
+ \sinh \theta_m (z_m) e^{-i\varphi (z_m)} H_{0}^{(1)}\left( z_{m}\right),
\label{F-function}
\\
G_m \left( z_{m}\right) & =
\cosh \theta_m (z_m) e^{-i(\delta_m (z_m)+\psi_m (z_m))} H_{0}^{(2)}\left( z_{m}\right)
+ \sinh \theta_m (z_m) e^{-i\psi (z_m)} H_{0}^{(1)}\left( z_{m}\right),
\label{G-function}
\\
\bar{F}_m \left( z_{m}\right) & = z_m 
\left[
\cosh \theta_m (z_m) e^{-i(\delta_m (z_m)+\varphi_m (z_m))} \dfrac{d H_{0}^{(2)}\left( z_{m}\right)}{dz_m}
+ 
\sinh \theta_m (z_m) e^{-i\varphi (z_m)} \dfrac{d H_{0}^{(1)}\left( z_{m}\right)}{dz_m}
\right],
\label{F-bar-function}
\\
\bar{G}_m \left( z_{m}\right) & =
 z_m 
\left[
\cosh \theta_m (z_m) e^{-i(\delta_m (z_m)+\psi_m (z_m))} \dfrac{d H_{0}^{(2)}\left( z_{m}\right)}{dz_m}
+ 
\sinh \theta_m (z_m) e^{-i\psi (z_m)} \dfrac{d H_{0}^{(1)}\left( z_{m}\right)}{dz_m}
\right].
\label{G-bar-function}
\end{align}
It is a straightforward exercise to show that the fields $x^{i}\left( t,\sigma \right)$ and $\pi^{i}\left( t,\sigma \right)$ are canonically conjugate to each other in the untangled oscillator representation
\begin{equation}
\left[
x^{k}\left( t,\sigma \right) ,
\pi^{l}\left( t,\sigma' \right)
 \right]
= i \delta^{kl} \delta(\sigma - \sigma').
\label{ETCR-ab}
\end{equation}
The identical copy of the system can be constructed by applying the tilde-conjugation given in the Appendix to the fields 
$x^{i}\left( t,\sigma \right)$ and $\pi^{i}\left( t,\sigma \right)$, respectively. We write down the result for completeness
\begin{align}
\tilde{x}^{i}\left( t,\sigma \right) & = x_{0}^{i}+p_{0}^{i}t
-\frac{i}{2}
\sqrt{\frac{\pi}{H}}
\sum _{m>0}\left[ \tilde{a}_{m}^{i}\left( z_{m}\right)e^{-im\sigma } F^{*}_m \left( z_{m}\right) + 
\tilde{a}_{m}^{i\dagger}\left( z_{m}\right) e^{im\sigma } F_{m} \left( z_{m}\right)\right]
\nonumber
\\
& +\dfrac {i} {2}\sqrt {\dfrac {\pi } {H}}
\sum _{m>0}\left[ \tilde{b}_{m}^{i}\left( z_{m}\right)e^{im\sigma } G^{*}_m \left( z_{m}\right) + 
\tilde{b}_{m}^{i\dagger}\left( z_{m}\right) e^{-im\sigma } G_{m} \left( z_{m}\right)\right]
,
\label{sol-x-tilde-ab}
\\
\tilde{\pi}^{i}\left( t,\sigma \right) & = \frac{1}{2\pi} p_{0}^{i}
-\frac{i}{2}
\sqrt{\frac{H}{\pi}}
\sum _{m>0}\left[ \tilde{a}_{m}^{i}\left( z_{m}\right)e^{-im\sigma } \bar{F}^{*}_m \left( z_{m}\right) + 
\tilde{a}_{m}^{i\dagger}\left( z_{m}\right) e^{im\sigma } \bar{F}_{m} \left( z_{m}\right)\right]
\nonumber
\\
& +\dfrac {i} {2}\sqrt {\dfrac {H } {\pi}}
\sum _{m>0}\left[ \tilde{b}_{m}^{i}\left( z_{m}\right)e^{im\sigma } \bar{G}^{*}_m \left( z_{m}\right) + 
\tilde{b}_{m}^{i\dagger}\left( z_{m}\right) e^{-im\sigma } \bar{G}_{m} \left( z_{m}\right)\right]
.
\label{sol-p-tilde-ab}
\end{align}
Then by combining the equations (\ref{sol-x-ab}), (\ref{sol-p-ab}), (\ref{sol-x-tilde-ab}) and (\ref{sol-p-tilde-ab}) we obtain the following thermal doublet fields
\begin{align}
\phi^{i\alpha}\left( t,\sigma \right) & = x_{0}^{i\alpha}+p_{0}^{i\alpha}t
+\frac{i}{2}
\sqrt{\frac{\pi}{H}}
\sum _{m>0}\left[ a_{m}^{i\alpha}\left( z_{m}\right)e^{im\sigma } F_m \left( z_{m}\right) - 
\left(s_3 \, \bar{a}_{m}^{i T}\left( z_{m}\right)\right)^{\alpha} e^{-im\sigma } 
F^{*}_{m} \left( z_{m}\right)\right]
\nonumber
\\
& + \dfrac {i} {2}\sqrt {\dfrac {\pi } {H}}
\sum _{m>0}\left[ b_{m}^{i\alpha}\left( z_{m}\right)e^{-im\sigma } G_m \left( z_{m}\right) -
\left(s_3 \, \bar{b}_{m}^{i T}\left( z_{m}\right)\right)^{\alpha} e^{im\sigma } G^{*}_{m} \left( z_{m}\right)\right]
,
\label{sol-phi-doublet-ab}
\\
\pi^{i\alpha}\left( t,\sigma \right) & = \frac{1}{2\pi} p_{0}^{i\alpha}
+\frac{i}{2}
\sqrt{\frac{H}{\pi}}
\sum _{m>0}\left[ a_{m}^{i\alpha}\left( z_{m}\right)e^{im\sigma } \bar{F}_m \left( z_{m}\right) - 
\left(s_3 \, \bar{a}_{m}^{i T}\left( z_{m}\right)\right)^{\alpha} e^{-im\sigma } \bar{F}^{*}_{m} \left( z_{m}\right)\right]
\nonumber
\\
& +\dfrac {i} {2}\sqrt {\dfrac {H } {\pi}}
\sum _{m>0}\left[ b_{m}^{i\alpha}\left( z_{m}\right)e^{-im\sigma } \bar{G}_m \left( z_{m}\right) + 
\left(s_3 \, \bar{b}_{m}^{i T}\left( z_{m}\right)\right)^{\alpha} e^{im\sigma } \bar{G}^{*}_{m} \left( z_{m}\right)\right]
.
\label{sol-pi-doublet-ab}
\end{align}
The physical interpretation of the fields $\phi^{i\alpha}\left( t,\sigma \right)$ and $\pi^{i\alpha}\left( t,\sigma \right)$ is the following: they represent the degrees of freedom associated to the two-dimensional $\sigma$-model that lie on the embedded cylinder $x(\Sigma^{1,1})$ in terms of which the nonequilibrium dynamics of the untangled modes is described in the cosmological gauge. From the point of view of $dS_4$, the theory is local in a neighbourhood of $x(\Sigma^{1,1})$ as the observation of the nonequilibrium dynamics is accessible only to the observer defined in the Section 2. 

The consistency of the present formulation of the nonequilibrium dynamics with the canonical quantization formalism requires the identification of the canonically conjugate field variables. They can be recognized from the set 
$\{ \phi^{i\alpha}\left( t,\sigma \right), \pi^{i\alpha}\left( t,\sigma \right), \bar{\phi}^{i\alpha}\left( t,\sigma \right), \bar{\pi}^{i\alpha}\left( t,\sigma \right) $ where the fields $\bar{\phi}^{i\alpha}\left( t,\sigma \right)$ and $\bar{\pi}^{i\alpha}\left( t,\sigma \right) $ can be obtained from $\phi^{i\alpha}\left( t,\sigma \right)$ and $\pi^{i\alpha}\left( t,\sigma \right) \}$, respectively, by applying the bar-conjugation operation given by the equation (\ref{bar-op}). After some algebra, one can prove that the following equal time commutation relations are satisfied
\begin{align}
\left[
\phi^{k\alpha}\left( t,\sigma \right) ,
\pi^{l\beta}\left( t,\sigma' \right)
 \right]
& = i \delta^{kl} s^{\alpha \beta}_{3} \delta(\sigma - \sigma'),
\label{ETCR-thermal-1}
\\
\left[
\bar{\phi}^{k\alpha}\left( t,\sigma \right) ,
\pi^{l\beta}\left( t,\sigma' \right)
 \right]
& = i \delta^{kl} \delta^{\alpha \beta} \delta(\sigma - \sigma'),
\label{ETCR-thermal-2}
\\
\left[
\phi^{k\alpha}\left( t,\sigma \right) ,
\bar{\pi}^{l\beta}\left( t,\sigma' \right)
 \right]
& = i \delta^{kl} \delta^{\alpha \beta} \delta(\sigma - \sigma'),
\label{ETCR-thermal-3}
\\
\left[
\bar{\phi}^{k\alpha}\left( t,\sigma \right) ,
\bar{\pi}^{l\beta}\left( t,\sigma' \right)
 \right]
& = i \delta^{kl} s^{\alpha \beta}_{3} \delta(\sigma - \sigma').
\label{ETCR-thermal-4}
\end{align}
The above equations show that the bar-conjugation is a necessary operation in the theory because the canonical conjugate variables can be obtained only by applying it to the original fields.

A generalization of the generator of the partial evolution from the thermal doublet oscillators to the thermal doublet fields is obtained by studying the time variation of the fields $\phi^{i\alpha}\left( t,\sigma \right)$, $\pi^{i\alpha}\left( t,\sigma \right)$, $\bar{\phi}^{i\alpha}\left( t,\sigma \right)$ and $\bar{\pi}^{i\alpha}\left( t,\sigma \right)$. For the simplicity of derivation, we discuss that in the classical theory. Then from the equations (\ref{sol-phi-doublet-ab}) and (\ref{sol-pi-doublet-ab}) and their bar-conjugates we derive the following relations
\begin{align}
\frac{d \phi^{i\alpha}\left( t,\sigma \right)}{d t} & =
\frac{ \partial \phi^{i\alpha}\left( t,\sigma \right)}{\partial t} 
+
\partial_{\sigma} \overrightarrow{\mathbf{D}}^{\alpha \beta} (t) \phi^{i\beta}\left( t,\sigma \right),
\label{time-evolt-phi}
\\
\frac{d \pi^{i\alpha}\left( t,\sigma \right)}{d t} & =
\frac{ \partial \pi^{i\alpha}\left( t,\sigma \right)}{\partial t} 
+
\partial_{\sigma} \overrightarrow{\mathbf{D}}^{\alpha \beta} (t) \pi^{i\beta}\left( t,\sigma \right),
\label{time-evolt-pi}
\\
\frac{d \bar{\phi}^{i\alpha}\left( t,\sigma \right)}{d t} & =
\frac{ \partial \bar{\phi}^{i\alpha}\left( t,\sigma \right)}{\partial t} 
+
\partial_{\sigma} \bar{\phi}^{i\beta}\left( t,\sigma \right) \overleftarrow{\mathbf{D}}^{\beta \alpha} (t) ,
\label{time-evolt-bar-phi}
\\
\frac{d \bar{\pi}^{i\alpha}\left( t,\sigma \right)}{d t} & =
\frac{ \partial \bar{\pi}^{i\alpha}\left( t,\sigma \right)}{\partial t} 
+
\partial_{\sigma} \bar{\pi}^{i\beta}\left( t,\sigma \right) \overleftarrow{\mathbf{D}}^{\beta \alpha} (t) .
\label{time-evolt-bar-pi}
\end{align}
Here, we have introduced the operator
\begin{equation}
\mathbf{D}^{\alpha \beta} (t) = i \left[ \mathbf{f} (t) \delta^{\alpha \beta} + i \mathbf{n}(t) T^{\alpha \beta}_0 \right] \mathbf{M}^{-1},
\label{D-operator}
\end{equation}  
where $\mathbf{f} (t)$,  $\mathbf{n}(t)$ and $\mathbf{M}$ have the eigenvalues $f_m (t)$, $n_{a,m}(t) + n_{b,m} (t)$ and $m$ when acting on the thermal oscillating modes, for all $m > 0$. The matrix $T_{0}$ is defined by the following relation
\begin{align}
T_0 & =
\begin{pmatrix}
1 & -1
\\
1 & -1
\end{pmatrix}.
\label{T0-matrix}
\end{align}
By inspecting the equations (\ref{time-evolt-phi}) - (\ref{time-evolt-bar-pi}) we observe that they can be interpreted as the
generalized Hamilton equations of the fields $\phi^{i\alpha}\left( t,\sigma \right)$, $\pi^{i\alpha}\left( t,\sigma \right)$, $\bar{\phi}^{i\alpha}\left( t,\sigma \right)$ and 
$\bar{\pi}^{i\alpha}\left( t,\sigma \right)$ generated by a Hamiltonian-like functional $\hat{\mathit{H}}_{tot}(t)$. 
Under the assumption that $\hat{\mathit{H}}_{tot}(t)$ does not contain derivatives of momenta with respect to ${\sigma}$ (supported by the equation (\ref{fixed-action-cosmo-gauge})) and after some simple calculations we obtain the following functional
\begin{align}
\hat{\mathit{H}}_{tot}(t) & = \frac{1}{2\pi} \int_{0}^{2 \pi} d \sigma \sum_{\kappa = a, b}
\left[
4 \pi \bar{\pi}^{i\alpha}_{\kappa}\left( t,\sigma \right)
\pi^{i\alpha}_{\kappa}\left( t,\sigma \right)
+ 2 e^{4Ht} 
\bar{\phi}^{i\alpha}_{\kappa} \left( t,\sigma \right) 
\partial^{2}_{\sigma} \phi^{i\alpha}_{\kappa} \left( t,\sigma \right)
\right.
\nonumber
\\
& +
\bar{\pi}^{i\alpha}_{\kappa}\left( t,\sigma \right)
\partial_{\sigma} 
\left( 
\overrightarrow{\mathbf{D}}^{\alpha \beta}_{\kappa} (t)
\phi^{i\beta}_{\kappa} \left( t,\sigma \right)
\right)
+
\partial_{\sigma} 
\left(
\bar{\phi}^{i\beta}_{\kappa} \left( t,\sigma \right)
\overleftarrow{\mathbf{D}}^{\beta \alpha}_{\kappa} (t)
\right)
\pi^{i\alpha}_{\kappa}\left( t,\sigma \right)
\nonumber
\\
& 
-
\left.
\bar{\phi}^{i\alpha}_{\kappa} \left( t,\sigma \right) 
\partial_{\sigma} 
\left( 
\overrightarrow{\mathbf{D}}^{\alpha \beta}_{\kappa} (t)
\pi^{i\beta}_{\kappa} \left( t,\sigma \right)
\right)
-
\partial_{\sigma}
\left(
\bar{\pi}^{i\beta}_{\kappa} \left( t,\sigma \right)
\overleftarrow{\mathbf{D}}^{\beta \alpha}_{\kappa} (t)
\right)
\phi^{i\alpha}_{\kappa}\left( t,\sigma \right)
\right]
\nonumber
\\
& -
\frac{e^{4Ht}}{\pi} \int_{0}^{2 \pi} d \sigma \sum_{\kappa = a, b}
\partial_{\sigma}
\left(
\bar{\phi}^{i\alpha}_{\kappa} \left( t,\sigma \right)
\partial_{\sigma} 
\phi^{i\alpha}_{\kappa} \left( t,\sigma \right)
\right)
.
\label{total-Hamiltonian-like}
\end{align} 
In the above relation, $\mathbf{D}^{\alpha \beta}_{a(b)} (t)$ are given by the equation (\ref{D-operator}) in which only $n_{a(b),m}(t)$ appear instead of the sum. The first line of (\ref{total-Hamiltonian-like}) is standard and describe the free dynamics of the thermal doublet fields. The second and the third lines contain the contribution of the Bogoliubov transformations which is expected to be modified by the interaction. In the standard NETFD theory, these terms are considered to be part of the free evolution, too, as they are compensated by a counterterm in the theory with interactions \cite{Umezawa:1993yq}. The last term is a total derivative that can be made to vanish by choosing proper boundary conditions on the thermal doublet fields in the variable $\sigma$, for example the periodic boundary conditions.

\section{Discussion}

In this paper, we have continued the analysis of the two-dimensional $\sigma$-model with the de Sitter target space  in the cosmological gauge started in \cite{Vancea:2016tkt}. We have addressed here two fundamental issues of this field theory: the problem of constructing the untangled oscillator representation of the $\sigma$-model and the thermalization of corresponding modes. Both frequencies and operators of the untangled oscillators that have been found are time-dependent. Their existence depends on finding consistent Bogoliubov transformations that can diagonalize the Hamiltonian. The consistency is expressed by the equations (\ref{theta-constraint}) and (\ref{reality-constraint}) that must have real solutions. The untangled modes obey partial evolution equations of the Heisenberg type (\ref{Heisenberg-equations}) that determine the local evolution of the untangled modes along the local integral line of the time-dependent Killing vector of the de Sitter space. Although untangled among themselves, the left and right moving modes still interact with the time-dependent background metric. This suggests that the thermalization is in nonequilibrium since the exchange of energy between the background and the oscillators takes place continuously. The thermalization of the untangled modes has been discussed in the NETFD formalism. This method allows one to calculate sensible quantities of the theory such as the local one-body propagator of an arbitrary thermal mode. Next, we have shown that the thermal doublet oscillators belong to a thermal doublet field with a canonical structure. We have determined the time variation of the thermal field variables and derived a Hamiltonian-like functional $\hat{\mathit{H}}_{tot}(t)$ for which this variation can be interpreted as generalized Hamilton equations. This functional can be straightforwardly generalize to an operator in the quantized theory. 

Compared with the literature, the equal time commutation relations (\ref{ETCR-thermal-1}) - (\ref{ETCR-thermal-4}) and the operator $\mathbf{D}^{\alpha \beta} (t)$ are similar to the ones obtained in \cite{Mizutani:2011rt} for the relativistic neutral scalar field in the Minkowski space-time. Also, the quantum and thermal field theories presented in this paper has close similarities with the little strings in an arbitrary time-dependent gauge \cite{Li:2007gf} and with the string theory near null cosmological singularities of space-time \cite{Nardi:2011sk,Madhu:2009jh,Narayan:2010rm}. The two-dimensional $\sigma$-model with the de Sitter target space in the cosmological gauge has turned to be a quite rich theory. It seems that most of this structure can be analysed locally by employing canonical methods which is a positive thing. A more detailed analysis of the canonical structure, further applications and generalizations of this model are currently under study.

\section*{Acknowledgements}
It is a pleasure to acknowledge R. J. Scherer Santos and  C. F. L. Godinho for discussions and to J. A. Helay\"{e}l-Neto for hospitality at Centro Brasileiro de Pesquisas F\'{i}sicas (CBPF) where part of this work was completed.

\section*{Appendix}

In this section we briefly review the set of axioms of the NETFD following \cite{Umezawa:1993yq}. 

\begin{enumerate}
\item The thermalization process of a physical system described by the (anti-) commuting field operator
 $\phi\left(  x \right)  $ is realized by doubling the original degrees of freedom
to $( \phi\left(  x\right), \tilde{\phi}\left(  x\right) )$. The original and tilde set of operators commute with eaach other and
act on the thermal or total Hilbert space which is the direct product of the two identical Hilbert spaces
\begin{equation}
\widehat{\mathcal{H}}=\mathcal{H}\otimes\mathcal{\tilde{H}}. 
\label{NETFD-Ax-1}%
\end{equation}
To every operator of the original system $O=O\left(  \phi\left(  x\right)  ,\phi^{\dag}\left(
x\right)  \right)  $ that act on $\widehat{\mathcal{H}}$ it must be associated an identical operator
acting on the tilde system $\tilde{O}=O^{\ast}\left(  \tilde{\phi}\left(  x\right)
,\tilde{\phi}^{\dag}\left(  x\right)  \right)  $ .

\item The tilde operation defines an involution on $\widehat{\mathcal{H}}$ that 
has the following properties
\begin{align}
\left(  c_{1}O_{1}+c_{2}O_{2}\right)  ^{\symbol{126}}  &  =c_{1}^{\ast}%
\tilde{O}_{1}+c_{2}^{\ast}\tilde{O}_{2},\label{Ax2-1}\\
\left(  O_{1}O_{2}\right)  ^{\symbol{126}}  &  =\tilde{O}_{1}\tilde{O}%
_{2},\label{Ax2-2}\\
\left(  O^{\dag}\right)  ^{\symbol{126}}  &  =\tilde{O}^{\dag},\label{Ax2-3}\\
\left(  \tilde{O}\right)  ^{\symbol{126}}  &  =\varepsilon O, \label{Ax2-4}%
\end{align}
for all $c_{1},c_{2}\in%
\mathbb{C}
$ and all $O,O_{1},O_{2}\in End(\widehat{\mathcal{H}})$. Here, $\varepsilon=+1(-1)$
for bosonic (fermionic) operators.

\item The vacuum state of the thermalized system is invariant under the tilde involution 
\begin{equation}
\widetilde{\left\vert 0(t)\right\rangle }=\left\vert 0(t)\right\rangle
,\qquad\widetilde{\left\langle 0(t)\right\vert }=\left\langle 0(t)\right\vert
. \label{Ax-3}%
\end{equation}

\item The generator of the time-evolution $\hat{H}$ satisfies the following
relation
\begin{equation}
\left(  i\hat{H}\right)  ^{\symbol{126}}=i\hat{H}. \label{Ax-4}%
\end{equation}
It follows that the Hamiltonian can be written as the difference between 
the original Hamiltonian and the tilde Hamiltonian as follows
\begin{equation}
\hat{H}=H-\tilde{H}. \label{Ax-4-a}%
\end{equation}

\item There is a set of operators $\left\{  \xi,\tilde{\xi},\xi^{\dag}%
,\tilde{\xi}^{\dag}\right\}  $ that define the time-independent free
quasi-particle representation of the total system. These operators act on the vacuum
as follows
\begin{equation}
\xi\left\vert 0\right\rangle =\tilde{\xi}\left\vert 0\right\rangle
=0,\qquad\left\langle 0\right\vert \xi^{\dag}=\left\langle 0\right\vert
\tilde{\xi}^{\dag}=0. \label{Ax-5}%
\end{equation}
The thermal states are defined by the thermal state condition%
\begin{align}
\left\langle 0\right\vert O(t)  &  =\left\langle 0\right\vert \tilde{O}^{\dag
}(t)\qquad\text{for bosons},\label{Ax-6-a}\\
\left\langle 0\right\vert O(t)  &  =e^{i\theta}\left\langle 0\right\vert
\tilde{O}^{\dag}(t)\qquad\text{for fermions}, \label{Ax-6-b}%
\end{align}
where $\theta$ is determined by the tilde conjugation rules.

\item The dynamical observables $O$ are defined in terms of non-tilde operators only. Their thermal average
is defined by its vacuum expectation value in the time-independent vacuum 
\begin{equation}
\left\langle O\right\rangle =\left\langle 0\left\vert O\right\vert
0\right\rangle . \label{Ax-7}%
\end{equation}

\item For any value of the time variable $t$, there is an invertible map
between the total Hilbert space and the time-dependent quasi-particle
representation given by a time-dependent Bogoliubov transformation%
\begin{equation}%
\begin{pmatrix}
\phi\left(  t\right) \\
e^{i\theta}\tilde{\phi}\left(  t\right)
\end{pmatrix}
=B^{-1}\left(  t\right)
\begin{pmatrix}
\xi\left(  t\right) \\
e^{i\theta}\tilde{\xi}\left(  t\right)
\end{pmatrix}
. \label{Ax-8}%
\end{equation}

\item The ket-vacuum and the bra-vacuum of the time-dependent quasi-particle representation
is defined by the following relation
\begin{align}
\xi\left(  t\right)  \left\vert 0(t)\right\rangle  &  =\tilde{\xi}\left(
t\right)  \left\vert 0(t)\right\rangle =0,\label{Ax-9-a1}\\
\left\langle 0(t)\right\vert \xi^{\dag}\left(  t\right)   &  =\left\langle
0(t)\right\vert \tilde{\xi}^{\dag}\left(  t\right)  =0. \label{Ax-9-b1}%
\end{align}
The time-evolution of the time-dependent bra- and ket-vacua is given by the
equations%
\begin{align}
i\frac{\partial}{\partial t}\left\vert 0(t)\right\rangle  &  =\hat
{H}\left\vert 0(t)\right\rangle ,\label{Ax-9-a2}\\
\left\langle 0(t)\right\vert \hat{H}  &  =0. \label{Ax-9-b2}%
\end{align}

\item The stationary thermal states in the Schroedinger representation are
defined by the equation%
\begin{equation}
\lim_{t\rightarrow\infty}\hat{H}\left\vert 0(t)\right\rangle =0,\label{Ax-10}%
\end{equation}
assuming that the thermal equilibrium is obtained at $t\rightarrow\infty$.
\end{enumerate}

Since the NETFD is a canonical formalism, the field operators can be decomposed in terms of
time-dependent oscillator operators $\left\{
a(t),a^{\dagger}(t),\tilde{a}(t),\tilde{a}^{\dagger}(t)\right\}  $. The family
of time-parametrized mappings between the canonical representation and the
time-dependent quasi-particle representations $\left\{  \xi\left(  t\right)
,\xi^{\dag}\left(  t\right)  ,\tilde{\xi}\left(  t\right)  ,\tilde{\xi}^{\dag
}\left(  t\right)  \right\}  $ is given by the time-dependent invertible
Bogoliubov map
\begin{equation}
B(t):\left\{  a(t),a^{\dagger}(t),\tilde{a}(t),\tilde{a}^{\dagger}(t)\right\}
\longrightarrow\left\{  \xi\left(  t\right)  ,\xi^{\dag}\left(  t\right)
,\tilde{\xi}\left(  t\right)  ,\tilde{\xi}^{\dag}\left(  t\right)  \right\}  .
\label{Ax-11}%
\end{equation}
The properties of a quasi-particle representation are specified by the properties of the 
Bogoliubov transformations (\ref{Ax-11}) that can satisfy any of the following
conditions: 

B.1) it can preserve the canonical structure:
\begin{equation}
B(t)\left(  s_{2}\otimes s_{3}\right)  B^{T}(t)  =s_{2}\otimes
s_{3},\label{Ax-12-a}
\end{equation}

B.2) it can preserve the Hermitian conjugation:
\begin{equation}
B^{\ast}(t)\left(  s_{1}\otimes\mathbf{1}\right)    =\left(  s_{1}%
\otimes\mathbf{1}\right)  B(t),\label{Ax-12-b}
\end{equation}

B.3) it can preserve the tilde conjugation operation:
\begin{equation}
B^{\ast}(t)\left(  s_{1}\otimes s_{1}\right)    =\left(  s_{1}\otimes
s_{1}\right)  B(t), \label{Ax-12-c}
\end{equation}
Here, we have denoted by $s_{i}, i = 1, 2, 3$, the Pauli matrices. The first
requirement is made in all cases. The last two ones define two non-standard
representations of the thermal field theory. 

The most general linear Bogoliubov operator in the
doublet representation given by the equation (\ref{a-double}) has the following form
\begin{equation}
B_{\mathbf{k}}(t)=%
\begin{pmatrix}
1+\varepsilon n_{\mathbf{k}}(t) & -n_{\mathbf{k}}(t)\\
-\varepsilon & 1
\end{pmatrix}
. \label{Ax-13}%
\end{equation}
Here, $\varepsilon=+1(-1)$ for bosons (fermions). The number density 
 $n_{\mathbf{k}}(t)$ is defined by the following relation
\begin{equation}
n_{\mathbf{k}}(t)\delta\left(  \mathbf{k-l}\right)  =\left\langle
0(t)\left\vert a_{\mathbf{k}}^{\dag}a_{\mathbf{l}}\right\vert
0(t)\right\rangle . \label{Ax-14}%
\end{equation}

\end{document}